\documentstyle[epsf,psfig,preprint,aps]{revtex}
\begin {document}
\draft
\title{Signature of the existence of the positronium molecule}
\author{J. Usukura$^1$, K. Varga$^{2,3}$,  and Y. Suzuki$^4$}
\address{
$^1$Graduate School of Science and Technology, Niigata University,
Niigata 950-2181, Japan
\\
$^2$Institute of Nuclear Research of the Hungarian Academy of Sciences 
(ATOMKI), Debrecen, H-4001, Hungary
\\
$^3$Institute for Physical and Chemical Research (RIKEN), Wako, Saitama
351-0198, Japan
\\
$^4$Department of Physics, Faculty of Science, Niigata University,
Niigata 950-2181, Japan}

\date{\today}

\maketitle

\begin{abstract}
The positronium molecule (Ps$_2$) has not been experimentally observed 
yet because its tiny (4.5 eV) binding energy cannot be detected when the
molecule annihilates by emitting two photons with energy of 0.51 MeV each. 
It is shown in this paper 
that the electric dipole transition between the recently found 
$L=1$ excited-state and the $L=0$ ground-state with its characteristic
photon energy of 4.94 eV is a clear signature of the 
existence of the positronium molecule and the possibility of
its experimental observation is realistic. The probability of this
transition is about 17 \% of the total decay rate. An other 
Coulomb four-body system containing positron, 
HPs (the positronium hydride or hydrogen positride), is 
also included for comparison.

\end{abstract}

\narrowtext

\section{Introduction}
Despite the early theoretical prediction of its existence \cite{Ore}, 
the Ps$_2$ molecule has not been  experimentally found
to date. The difficulty stems from the fact that this system is neutral
and therefore it cannot be separated from the positronium atoms (Ps) and
its primary decay mode, the annihilation by two-photon emission,
is exactly the same as that of the Ps atom. The energy of the photons
arising from the annihilation is different in principle: The photons
carry 1.02 MeV energy due to the annihilation plus the binding energy
of the corresponding system. The binding-energy difference is, however, less
than 1 eV and adding it to 1.02 MeV, the energy of the photons
coming from the Ps atom or Ps$_2$ molecule cannot be experimentally
distinguished. The experimental observation of the biexcitons
can be considered as an indirect indication of the existence of Ps$_2$.
\par\indent
In our recent Letter \cite{vus} we have predicted the existence of
a hitherto unknown bound excited-state of the Ps$_2$ molecule.
In this paper we give a detailed description of this state. We have
investigated possible decay modes of this state with a special
emphasis on the electric dipole ($E1$) transition to the ground state. 
It will be shown
that the probability of the $E1$ transition is comparable to
that of the annihilation. The unique energy of this transition
may possibly be utilized as a sign for the experimental identification of the
Ps$_2$ molecule.
\par\indent
The stochastic variational method  
\cite{SVM,prca} has been used to solve the Coulomb four-body problem. 
In this method the variational
trial functions are optimized by gambling: Randomly chosen configurations
are probed and most adequate functions are selected to be
the basis states. 
\par\indent
The Correlated Gaussians (CG) \cite{cg1} are used as basis
functions in this procedure. The CG basis has a long history in
atomic and molecular physics and highly accurate calculations are based 
on this form of basis functions \cite{vus,cg2,suv,posi,posi1,posi2}.
The angular part is given by the global vector representation 
\cite{suv}. This approach greatly simplifies the calculations
for non-spherical systems by replacing the partial wave expansion
with a much simpler representation of the angular motion.
\par\indent
The hydrogen positride (positronium hydride), HPs, 
has already been in the focus of intensive theoretical and 
experimental investigation. This is an ideal system to test the SVM. 
We compare the properties of the Ps$_2$ and HPs molecules.
\par\indent
The plan of this paper is as follows. In sect. II we give
a brief description of the trial function and the stochastic variational
method. In sect. III the results are presented. The main results of the
paper are summarized in sect. IV. In Appendices A-D we collect 
some basic ingredients which are used in the present study in order 
to help readers reproduce our results: formulae 
of the matrix elements in the CG basis, the separation of the 
center-of-mass motion from the CG basis, the use of Sherman-Morrison 
formula in selecting nonlinear parameters, and the symmetry requirement 
for the trial wave function of the Ps$_2$ molecule.

\section{The calculation}
A system of two electrons with mass $m$ and two positive unit charges
of mass $M$ is considered. Their relative mass is characterized
by the ratio $\sigma=m/M$, and the positronic limit is realized by
$\sigma=1$. (Though we consider the case of $\sigma=1$ in this paper, 
the extension to other $\sigma$ values is straightforward, so we 
give a formulation assuming an arbitrary mass ratio.) 
The Hamiltonian of the system reads as
\begin{equation}
H=\sum_{i=1}^4 T_i - T_{\rm cm}+\sum_{i<j} {q_i q_j \over {\vert 
{\bf r}_i-{\bf r}_j \vert }} ,
\end{equation}
where $q_i$ and ${\bf r}_i$ are the charges and the position 
vectors of the particles.
Particle labels 1 and 3 denote the positive charges, while labels 
2 and 4 denote the negative charges. 
A relative coordinate system is introduced
by defining ${\bf x}_1$ and ${\bf x}_2$ as the distance vectors
between the positive and negative charges in the first and second 
atom, and ${\bf x}_3$ as the distance vector between the center-of-masses 
of the two atoms:
\begin{eqnarray}
{\bf x}_1 & = & {\bf r}_1 - {\bf r}_2, \label{eq:x1}\\
{\bf x}_2 & = & {\bf r}_3 - {\bf r}_4, \\
{\bf x}_3 & = & \frac{M{\bf r}_1+m{\bf r}_2}{M+m}-
\frac{M{\bf r}_3+m{\bf r}_4}{M+m}, \\
{\bf x}_4 & = & {\bf R} =  
\frac{M{\bf r}_1+m{\bf r}_2+M{\bf r}_3+m{\bf r}_4}{2M+2m}.
\label{eq:x4}
\end{eqnarray} 
We use the abbreviation ${\bf x}=\lbrace{\bf x}_1,....,{\bf x}_4\rbrace$
and ${\bf r}=\lbrace{\bf r}_1,...,{\bf r}_4\rbrace$.

\subsection{The wave function}
The CG of the form
\begin{equation}
G_{A}({\bf r})={\rm exp} \lbrace -{1\over 2} \tilde{{\bf r}} 
A {\bf r}\rbrace =
{\rm exp} \lbrace -{1\over 2} \sum_{i,j=1}^4
A_{ij}{\bf r}_i \cdot {\bf r}_j\rbrace
\end{equation}
is very popular in atomic and molecular physics 
\cite{prca,cg1,cg2,suv,posi,posi1,posi2}. Here $\tilde{\bf r}$ 
stands for a one-row vector whose $i$th element is ${\bf r}_i$. 
The merit of this basis is that  the matrix elements 
are analytically available and unlike other trial functions 
(for example, Hylleraas-type  functions) one can relatively easily 
extend the basis for the case of more than three particles. The well-known 
defects of this basis are that it does not fulfill the cusp condition
and its asymptotics does not follow the exponential falloff. This latter
problem, especially for bound states, can be cured by taking 
linear combinations of adequately chosen CGs. 

\par\indent
The CG defined above is spherical and can thus describe systems with only 
$L=0$ orbital angular momentum. The usual way to account for the orbital
motion in the case of $L\ne 0$ is the partial-wave
expansion. Because of the complexities arising from the evaluation of 
matrix elements this expansion gets very tedious for more than
three particles. To avoid this difficulty the global vector representation
\cite{suv} is used. In this approach, one defines a vector
${\bf v}$ as a linear combination of the relative coordinates:
\begin{equation}
{\bf v}=\sum_{i=1}^{4}u_{i} {\bf r}_i ,
\end{equation}
and the non-spherical part of the wave function is represented by a 
solid spherical harmonic
\begin{equation}
{\cal Y}_{KLM}({\bf v})=v^{2K+L}Y_{LM}({\hat {\bf v}}) .
\end{equation}
The linear combination coefficients $u_i$ are considered to be 
variational parameters and their optimal values are to be determined
by the SVM as will be discussed later. The details and examples can 
be found in \cite{suv}.
\par\indent
The calculation of the matrix elements for the space part of 
our basis function
\begin{equation}
\label{fklm}
f_{KLM}(u,A,{\bf r})=G_A({\bf r}) {\cal Y}_{KLM}({\bf v})
\end{equation}
is given in \cite{suv}. 
In the special case of $K=0$ the matrix 
elements can be written in much simpler form. This is shown in 
Appendix A. In the $K\ne 0$ case, the CG is multiplied by a polynomial
of the relative coordinates. In some cases this might be very useful, 
it can improve the short-distance behavior, for example, but this role can also
be played by an appropriate superposition of the exponentials. We use $K=0$
in this paper.

\par\indent
The translational invariance of the wave function is ensured by requiring 
that the parameters $A$ and $u$ fulfill some special conditions. 
As is detailed in Appendix B, these conditions  ensure that the 
motion of the center-of-mass is factorized in a product form. 

\par\indent
By combining the CG with the angular and spin parts, the full basis function
takes the form
\begin{equation}
\Phi_{kLS}={\cal A}\lbrace \chi_{SM_S} f_{KLM}(u_k,A_k,{\bf r}) \rbrace , 
\end{equation}
with an appropriate spin function $\chi_{SM_S}$, where  
``$k$'' is the index of the basis states and ${\cal A}$ 
is an antisymmetrizer for the identical fermions. In the positronium 
limit $(\sigma=1)$ the Hamiltonian becomes invariant with respect to 
the interchange of positive and negative charges. Therefore 
the basis function should have a definite parity under the 
charge-permutation operator. See Appendix D for the details of the symmetry 
requirement on the wave function. 
For the special case with $S=0$ and $M_S=0$ in which two spins of 
positive charges and 
two electron spins are coupled to zero, respectively, the spin part of 
the wave function reads as
\begin{equation}
\label{spinzero}
\chi_{00}={1 \over 2} \Big( \vert \uparrow \uparrow \downarrow 
\downarrow \rangle
-\vert \uparrow \downarrow \downarrow \uparrow \rangle 
-\vert \downarrow \uparrow \uparrow \downarrow \rangle 
+\vert \downarrow \downarrow \uparrow \uparrow \rangle 
\Big) .
\end{equation}
(Note that particles 1 and 3 are positive unit charges 
and particles 2 and 4 are electrons.)
\par\indent
Instead of optimizing the parameters of $A$ it is more advantageous 
to rewrite Eq. (6) as
\begin{equation}
{\rm exp}\Big\lbrace -{1\over 2} \sum_{i<j} \alpha_{ij} 
({\bf r}_i-{\bf r}_j)^2-{1\over 2}\sum_{i} \beta_i r_i^2 \Big\rbrace.
\label{basis}
\end{equation}
The relationship between ${\alpha}_{ij},{\beta}_i$ and $A$ is
\begin{equation}
{\alpha}_{ij}=-A_{ij} \ \ \ \ (i\ne j),  \ \ \ \ 
{\beta}_i=\sum_{k} A_{ki},
\label{apara}
\end{equation}
where $\alpha_{ji}\, (i<j)$ is assumed to be equal to $\alpha_{ij}$.
There are two reasons to choose this form. The first is that 
in choosing $\alpha_{ij}$ in this way we deal with a correlation function
between the particles $i$ and $j$, while $A_{ij}$ has no such direct 
meaning and during the optimization it is more difficult to limit 
the numerical interval of $A_{ij}$ to be chosen from. Secondly, one can
utilize this specific form to make the individual steps of the
parameter selection very fast. By taking a look at the expressions 
of the matrix elements in Appendix A, it is clear that the main
computational load is the calculation of the inverse and determinant
of the matrix of the nonlinear parameters. The form in Eq. (\ref{eq:sher})
offers the possibility of the usage of the Sherman-Morrison 
formula to calculate these quantities, leading to a much faster function
evaluation. The details of this step are given in Appendix C. 

\subsection{Electric dipole transition rate}
In the positronium  limit $(\sigma=1)$ we deal with antiparticles
and the electron-positron pair can annihilate.
The lifetime of the first excited-state with $L=1$ and negative parity 
is determined by both processes of 
annihilation and electric dipole transition to the ground state. 
The width $\Gamma_{\rm dipole}$ for the latter 
process is calculated through the 
reduced transition probability $B(E1)$ for the electric dipole operator
$D_{\mu}=\sum_{k=1}^4 q_k \vert {\bf r}_k-{\bf R} \vert 
Y_{1 \mu}(\widehat{{\bf r}_k-{\bf R}})\ (\mu=-1,\,0,\,1)$
\begin{equation}
\Gamma_{\rm dipole} ={16 \pi \over 9} 
\Big( {E \over \hbar c}\Big)^3 B(E1;1^{-}\to 0^{+}),  
\end{equation}
with
\begin{equation}
B(E1;1^{-}\to 0^{+}) = \sum_{\mu} 
\vert \langle 00\vert D_{\mu} \vert 1M \rangle \vert^2 ,
\end{equation}
where $E$ is the excitation energy of the first excited state.  

\subsection{Annihilation rate}
The most dominant annihilation of the first excited-state of Ps$_2$ 
is accompanied by the emission of 
two photons with energy of about 0.5 MeV each. The decay width 
$\Gamma_{2\gamma}$ for the 
annihilation can be estimated through the decay width 
$\Gamma_{2\gamma}^{\rm Ps}$ 
of the para-positronium in spin-singlet state. This decay width has to be 
multiplied by the number $N_0$ of positron-electron pairs which are 
in spin-singlet state in the Ps$_2$. In the Ps$_2$ excited state 
we have four positron-electron pairs, among which the probability that 
the pair is in spin-singlet state is 1/4 because the total spin of 
the first excited-state of Ps$_2$ 
is zero, as will be shown later. ($N_0=4\times (1/4)=1$.) 
Therefore, to have an 
estimate for the decay due to the annihilation we can use the formula (2) 
of \cite{posi2}: 
  
\begin{equation}
\Gamma_{2\gamma}=N_0\Gamma_{2\gamma}^{\rm Ps}
\end{equation}
with
\begin{equation}
\Gamma_{2\gamma}^{\rm Ps}=4\pi \Big( {e^2 \over m c^2}\Big)^2\hbar c
\langle \Psi |\delta({\bf r}_1-{\bf r}_2)|\Psi \rangle
=4\pi \Big( {e^2 \over \hbar c}\Big)^4 \hbar c a_0^{-1}
\langle \delta(r_{12}) \rangle,
\end{equation}
where the probability of finding an electron at the position of a positron, 
$\langle \delta(r_{12}) \rangle$, is the expectation value of 
$\delta({\bf x}_1)$ given in a.u., that is $\langle \delta(r_{12}) \rangle$ 
is equal to $a_0^3 \langle \Psi |
\delta({\bf r}_1-{\bf r}_2)|\Psi \rangle $ with the Bohr radius $a_0$. 
Roughly speaking, the lifetime is inversely 
proportional to the probability of finding an electron and a 
positron at the same position. 
 
\subsection{The stochastic variational method}
To obtain very precise energy, one has to optimize the variational 
parameters 
$u_{ki}$ and $A_{kij}$ of the trial function.
The dimension of basis sets is typically between 100 and 1000,
and each basis state has nine nonlinear parameters. (See Appendix B.) 
The optimization of
a function with a few thousands nonlinear parameters cannot be
done efficiently by using a deterministic optimization method, since this 
could entail the complete reconstruction of the Hamiltonian matrix and 
diagonalization every time when some of the nonlinear parameters are altered.
Moreover, the deterministic search for the optimal value of  such a 
large number of parameters is likely to get trapped in a local minimum. 
\par\indent
A procedure based on the stochastic search for the best set of nonlinear
parameters can be programmed efficiently \cite{prca,cc} 
and is capable of achieving highly accurate results for most 
few-body systems \cite{vus,prca,suv,rmv}. The essence of the strategy can 
be summarized as follows: Let $\lbrace u_i,A_i \rbrace$ be the nonlinear 
parameters 
of the $i$th basis function out of the set of $K$ such basis functions. 
Then the procedure is
\vskip 0.3 cm
\par\noindent
(1) A succession of different sets of $(\lbrace u_i^{1},A_i^{1} \rbrace,...,
\lbrace u_i^{n_s},A_i^{n_s} \rbrace)$ are generated randomly.
\par\noindent
(2) By solving the eigenvalue problem, the corresponding energies 
$(E^1_i,...,E_i^{n_s})$ are determined.
\par\noindent
(3) The parameter set $\lbrace u_i^{m},A_i^{m} \rbrace$ 
which produces the lowest 
energy is then used to replace the existing $\lbrace u_{i},A_{i} \rbrace$ set.
\par\noindent
(4) The procedure cycles through the different parameter sets 
($\lbrace u_{i},A_{i} \rbrace,i=1,...,K)$, successively choosing different 
sets to minimize the energy until convergence is reached.
\vskip 0.3cm
\par\noindent
The essential reason motivating this strategy is the need to sample 
different sets of nonlinear parameters as fast as possible. The main
advantage is that  it is not necessary to recompute the complete
Hamiltonian nor it is necessary to solve the generalized eigenvalue problem 
from scratch each time a new parameter set is generated. By changing the 
elements of parameter set for each basis function individually, it is 
necessary to recompute only one row (column) of the Hamiltonian and 
overlap matrices each time the parameter set $\lbrace u_{i},A_{i} \rbrace$
is changed. Furthermore, the solution of the generalized eigenvalue 
problem is also expedited since the Hamiltonian matrix is already diagonal
apart from one row and one column. 
\par\indent
A similar strategy to the above was used when adding additional terms to 
the basis. 
\par\indent
The speed of the calculation can be further increased if one changes the 
nonlinear parameters $A_{i}$ in a special way. This is described 
in Appendix C. 
\par\indent
The above way of finding the best  parameters is certainly very restricted.
Even this simple method gives very accurate energies. More sophisticated 
technique may give better results in a smaller basis size.

\section{Results}
\par\indent
The results of calculations for the ground state of HPs
and Ps$_2$ and the first excited-state of Ps$_2$ are reported in this 
section.  The ground 
states of HPs and Ps$_2$ have already been subject to 
intensive calculations and some of the results obtained before for 
these systems serve as validation of the SVM. The calculation
of the properties of the excited state of the Ps$_2$ is the primary focus
of this paper. We have previously reported the energy of the ground state
of the Ps$_2$ and predicted the existence of an excited state of this
molecule. This paper reports considerably improved energies 
by further optimization of the nonlinear parameters of the basis.
The further optimization and the increase of the basis dimension
has produced an improved wave function and 
we present different properties of these systems by using 
that wave function. We show the convergence of the binding energies and 
various expectation values as a function of the dimension of the basis.
The results in the tables are shown for 
the basis dimensions of $K=100,200,400,800,1200,1600$. The basis has been 
subject of intensive optimizations at these dimensions. Once the 
optimization at a given basis size has been finished, new basis states 
have been added (each of them has been selected amongst hundreds of random
candidates) to reach the next basis size where the optimization is started 
again.  
While the pattern of convergence is a very useful information about the
accuracy of the results, one has to keep in mind that this can be distorted 
by many extraneous factors. This is because 
one cannot guarantee that the quality of these optimizations 
is the same. We expect that the stochastic selection of the basis is 
close to be the optimal choice
for lower dimensions, but for large dimensions ($K=1200,1600$) the 
procedure becomes more time consuming and we have less chance to find
the optimal parameters. 

\subsection{Hydrogen Positride, HPs}
The boundness of the exotic molecule, HPs, has been known theoretically 
for many years \cite{Ore2} and it has recently been created and 
observed in collisions between positrons and methane \cite{schrad}. 
The investigation of the stability of positronic atoms has been attracting 
much attention because positrons can be used as a tool for 
positron-annihilation spectroscopy in condensed matter physics. The 
HPs molecule is the simplest but ideal hydride to test the SVM. 
It is also very intriguing to see the difference between the properties
of Ps$_2$ and HPs. 

The energy calculated by SVM and by other methods
are shown in Table I. The proton mass is taken to be infinite. The two 
electrons are assumed to be in spin-singlet state. The spin states of the 
proton and the positron can be taken arbitrary. 
Our result, already at the dimension of $K=200$,
is better than the previous calculations. The increase of the basis size 
improves the energy further. The need of improved accuracy can be 
clearly seen in Table II, where various expectation values are listed.
The expectation value $\langle r^4_{e^-e^-}\rangle$, for example, is 
much less accurate than the energy and it is considerably improved beyond
the dimension $K=200$. 
\par\indent
One can compare the expectation values of the separation distances 
of the particles 
in the HPs to those in the H and Ps atoms. The average electron-positron 
distance is 3.48 $a_0$ in HPs, which is slightly different from that 
in the positronium atom (3 $a_0$). The average electron-proton distance 
in HPs and H is considerably different (2.31 $a_0$ and 1.5 $a_0$).
The average distance between the two positive charges (3.66 $a_0$)
is much larger than that in the H$_2$ molecule (1.41 $a_0$).

The correlation function defined by 
\begin{equation}
C({\bf r})=\langle \Psi | \delta({\bf r}_i-{\bf r}_j-{\bf r})
| \Psi \rangle
\end{equation}
 gives more detailed information on a system 
than just various average distances. This quantity can be calculated 
by using Eqs. (34) and (35). For the spherical wave function with $L=0$, 
$C({\bf r})$ is a function of $r$, that is, the monopole density, 
and for the $L=1$ wave function, it consists of two terms of monopole and 
quadrupole densities. Figure 1 displays $r^2\, C({\bf r})$ for various 
pairs of the constituents of HPs. The two electrons are 
attracted by the proton, 
but the proton-electron correlation function is much broader 
than that in the H atom, 
while they are separated with its maximum density being at about 2.8 a.u. 
The positron moves furthest from the proton and has a peak density around 
at 2.6 a.u. from the electron.

\par\indent
The $2\gamma$ annihilation rate, calculated from 
Eq. (16) with $N_0=2\times (1/4)=1/2$ and 
$\langle \delta_{e^+e^-}\rangle$ of Table II, is 
found to be $\Gamma_{2\gamma}=2.4722 \,{\rm ns}^{-1}$, 
improving the previous estimates \cite{frpsh} by about $0.5\%$.

\subsection{Positronium molecule, Ps$_2$: Ground state}
The energies by SVM are compared to the best previous results
in Table III. The result of SVM, again, already at the dimension of $K=200$,
is better than the energy of the previous calculations. 
The increase of the basis size improves the accuracy and the virial factor
$\vert 1+\langle V\rangle/(2\langle T\rangle)\vert$ becomes  as small as
$0.3\times 10^{-9}$, improving the 
previously best calculation by more than 4 order of magnitude. 
\par\indent
The average electron-positron distance is 4.487 $a_0$, which is about 1.5 
times larger than in the positronium atom.
The $2\gamma$ annihilation rate calculated from 
Eq. (16) by using $\langle \delta_{e^+e^-}\rangle$ of Table II is 
found to be $\Gamma_{2\gamma}=4.470 \,{\rm ns}^{-1}$.
\par\indent
The electron-electron and the electron-positron correlation functions
are compared in Fig. 2. The peak position of the electron-electron 
correlation function is 
shifted to larger distances compared to the one of the electron-positron 
correlation function. 
The electron-positron correlation function in Ps$_2$ has much broader 
distribution than the corresponding function in a Ps atom. 

\subsection{Positronium molecule, Ps$_2$: First excited state}
In our previous paper we have predicted the existence of the first
excited-state of the Ps$_2$ molecule. This is a unique bound-state which 
cannot decay into two Ps atoms 
due to the Pauli principle. The spin of this state is $S=0$ and the
orbital angular momentum is $L=1$ with negative parity. 
In this spin state, the Ps$_2$ molecule can dissociate into two
Ps atoms (bosons) only if the relative orbital angular momentum is even.
Consequently, the Ps$_2$ molecule with $L=1$ and negative parity 
cannot decay into the ground states of two Ps atoms
(Ps($L=0$)+Ps($L=0$)). The energy of this Ps$_2\,(L=1)$ state
($E=-$0.334408 a.u., see Table III) is lower than the energy of the 
relevant threshold ($-$0.3125 a.u.), and 
this state is therefore stable against the autodissociation 
into Ps($L=0$)+Ps($L=1$). The binding energy of this state is  
0.5961 eV, which is by about 40\% more than that of the ground state of
Ps$_2$ (0.4355 eV). 

\par\indent
We have shown in \cite{vus} that the bound excited state is essentially 
a system where two Ps atoms, one in its ground state and the other 
in its first excited $P$ state, are weakly coupled. The expectation 
value of the average 
electron-positron distance shown in Table V supports this picture: The 
value of 7.57 $a_0$ in the excited state is 15 \% 
larger than the average (6.5 
$a_0$) of the electron-positron distances in  
the $L=0$ ground state of the Ps atom (3 $a_0$) and the
$L=1$ excited state of the Ps atom (10 $a_0$). 
We can also estimate the root-mean-square distance 
$d=\sqrt{\langle {\bf x}_3\cdot{\bf x}_3 \rangle }$ between the two atoms by

\begin{equation}
d^2=\left\langle \left(\frac{{\bf r}_1+{\bf r}_2}{2}-
\frac{{\bf r}_3+{\bf r}_4}{2}\right)^2 \right\rangle=
\frac{1}{4}\Big( 2\langle r_{12}^2\rangle+\langle r_{13}^2\rangle
-2\langle {\bf r}_{12}\cdot{\bf r}_{14} \rangle\Big).
\end{equation}
The symmetry properties of the Ps$_2$ wave function are used to 
obtain the second equality. Using the values of Tables IV and V 
yields $d=6.93$ a.u. for the $L=1$ excited state and $d=4.82$ a.u. for the 
$L=0$ ground state. 

Figure 3 displays the 
electron-electron and electron-positron correlation functions. 
As mentioned before, the correlation function for the $L=1$ state 
consists of the monopole and quadrupole densities and their shapes 
depend on the magnetic quantum number $M$ of the wave 
function. Of course the $M$-dependence of the shapes 
is not independent of each other but is determined by the Clebsch-Gordan 
coefficient. See Eq. (34). The quadrupole density is contributed 
only from the $P$-wave for the electron-positron relative motion, while 
the monopole density is contributed by both $S$- and $P$-waves. 
Figure 3(a) plots the correlation functions 
for $M=0$ and Fig. 3(b) the correlation functions for $M=1$. 
As the correlation function is axially symmetric around the $z$ axis and 
has a reflection symmetry with respect to the $xy$ plane, 
the correlation function sliced on the $xz$ plane is drawn as a function of 
$x\,(x\ge 0),\, z\,(z\ge0)$. The electron-electron correlation function 
has a peak at the point corresponding to the average distance of 7.57 a.u. 
The electron-positron correlation function 
has two peaks reflecting the fact that the basic structure 
of the second bound-state is a weakly coupled system of a Ps 
atom in the $L=0$ state and another Ps atom in the $L=1$ spatially 
extended state. The peak located at a larger distance from the origin 
is due to the $P$-wave component of the Ps$_2$ molecule.

\par\indent
By using the obtained value for $\langle \delta(r_{12})\rangle$ in Eq. (16), 
the lifetime due to the annihilation is estimated to be 0.44 ns. 
This is about twice of the lifetime of the Ps$_2$ ground state. 
The $B(E1)$ value is calculated to be $B(E1)=0.87 e^2a_0^2$. By combining 
this value with the dipole transition energy of 4.94 eV, the 
lifetime due to the electric dipole transition has been found to be 2.1 ns. 
The branching of the electric dipole transition is thus about 17 \% of the 
total decay rate. Therefore, both branches contribute 
to the decay of the excited state of the Ps$_2$ molecule. 
Its lifetime is finally estimated to 
be about 0.37 ns. The excitation energy of 4.94 eV found for the Ps$_2$ 
is different by 0.16 eV 
from the corresponding excitation energy (5.10 eV) of a Ps atom. 
This difference seems to be large enough to detect its existence, e.g. 
in the photon absorption spectrum of the positronium gas. 

\section{Summary}
We have used the Correlated Gaussians combined with the angular functions 
which are specified by the global vector. Nonlinear parameters of the 
bases have been determined by the stochastic variational method. 
We have considerably improved the results of the previous calculations
for the ground state of HPs and Ps$_2$. In addition, 
we have calculated various
expectation values, correlation functions and other properties of the
excited state of the Ps$_2$ molecule. 

The excited state of the Ps$_2$ molecule has the orbital angular 
momentum $L=1$, the spin $S=0$, and negative parity. The excitation 
energy of the state is 4.941 eV, 0.596 eV below the threshold of 
Ps$(L=0)$+Ps$(L=1)$. Though this state is in the continuum of  
Ps$(L=0)$+Ps$(L=0)$ channel, it is stable against autodissociation 
into that channel because of the Pauli principle. 
The main decay mode of this state is the annihilation emitting 
two photons of about 0.5 MeV each which, except for the 
tiny binding-energy difference, is equal to the photon energies of Ps 
atoms. The annihilation decay mode is not useful to 
confirm experimentally the existence of the Ps$_2$ molecule. 
We have discussed a unique decay mode of the excited state, 
the electric dipole transition to the 
ground state. The lifetime due to the electric dipole transition has 
been calculated to be 2.1 ns, while the lifetime due to the annihilation is 
0.44 ns. The electric dipole transition can be used as a signal for 
experimental confirmation of the Ps$_2$ molecule.   

\bigskip
\par\indent
This work was supported by OTKA grant No. T17298 (Hungary) and 
Grant-in-Aid for International Scientific Research 
(Joint Research) (No. 08044065) of the Ministry of Education, Science 
and Culture (Japan). The authors are 
grateful for the use of RIKEN's computer facility which made possible 
most of the calculations. 

\section*{Appendix A:  Evaluation of matrix elements}
In this appendix the matrix elements of the spatial part of the
basis functions are given. The method of calculation of these 
analytical expressions is detailed in Refs. \cite{suv,book}. The 
main aim of this section is to convince the reader that the 
formulae are particularly simple for the case of $K=0$. The extension 
to a general $N$-body system is straightforward so that we assume that 
the system contains $N$ particles. 

\par\indent
The basic idea of the calculation of the matrix elements 
is the usage of the generating function $g$:

\begin{equation}
g({\bf s}; A, {\bf r})=\exp\Big(-{1\over 2}{\tilde{\bf r}}A{\bf r}+
\tilde{\bf s}{\bf r}\Big).
\end{equation}
In the special case of $K=0$, Eq. (\ref{fklm}) is obtained from $g$ by 

\begin{eqnarray}
\label{gfn}
f_{0LM}(u,A,{\bf r})&\equiv& {\rm e}^{-{1\over 2}{\tilde{{\bf r}}} A {\bf r}}
{\cal Y}_{0LM}({\bf v}) \nonumber \\
& =& {B_{L}\over L!}\int Y_{LM}({\hat {\bf e}}) 
\Bigg( {d^{L} \over d\lambda^{L}}g(\lambda{\bf e}u; A,{\bf r})
\Bigg)_{\lambda=0, e=1}d{\hat {\bf e}}, 
\end{eqnarray}
with
\begin{equation}
B_L={(2L+1)!!\over 4 \pi}.
\end{equation}

\par\indent
To abbreviate the expression for the matrix elements we introduce the 
following notation
\begin{equation}
\label{meofop}
\langle f'\vert {\cal O}\vert f\rangle=
\langle f_{0LM}(u',A',{\bf r})\vert
 {\cal O}\vert f_{0LM}(u,A,{\bf r})\rangle,
\end{equation}
where ${\cal O}$ stands for the unity, kinetic or potential energy
operators. The operators considered here are rotational invariant and 
thus the matrix elements are diagonal in $LM$. Note the prime on $f$
which is a reminder that the parameters in the ket and the bra
may be different.

\par\indent
The use of Eq. (\ref{gfn}) in Eq. (\ref{meofop}) leads to 
an expression that the
matrix element is derived from that between the generating functions, 
which becomes a function of 
parameters $\lambda, {\bf e}, \lambda'$ and ${\bf e}'$. Here the 
matrix element between the generating functions can be obtained 
easily by using the expression
\begin{equation}
\int  {\rm e}^{-{1\over 2}{\tilde{{\bf r}}}A {\bf r}+
{\tilde{{\bf s}}}{\bf r}}d{\bf r}
=\left({(2\pi)^{N} \over {\rm det} A }\right)^{3\over 2}
{\rm e}^{{1\over 2}{\tilde{{\bf s}}} A^{-1} {\bf s}} 
\label{gaussint1}
\end{equation}
and its extended formulae. 
After a power series expansion of the 
matrix element between the generating functions in terms of 
$\lambda, {\bf e}, \lambda'$ and ${\bf e}'$, 
the derivative and the integration prescribed in 
Eq. (\ref{gfn}) can be carried out straightforwardly \cite{suv,book}.

\par\indent
The overlap of the trial functions is given by
\begin{equation}
\langle f'\vert f\rangle=
\left({(2\pi)^{N} \over {\rm det}B}\right)^{3 \over 2} B_L \rho^L.
\end{equation}
The kinetic energy is expressed by
\begin{equation}
\langle f'\vert T-T_{\rm cm}\vert f\rangle= \frac{\hbar^2}{2}
\left(R  + LQ\rho^{-1}\right) \langle f'\vert f\rangle .
\end{equation}
The matrix elements of a central potential reads as
\begin{equation}
\langle f'\vert V(\vert {\bf r}_i-{\bf r}_j \vert) 
\vert f \rangle= \langle f'\vert f\rangle
\sum_{n=0}^{L} I(c,n) {L!\over (L-n)!} 
\Big({\gamma\gamma' \over c \rho}\Big)^{n} ,
\end{equation}
where the integral over the radial form of the potential is 
expressed with use of Hermite polynomials 
\begin{equation}
I(c,n)=
\frac{1}{\sqrt{\pi}(2n+1)!}  
\int_0^{\infty}V\Big(\sqrt{{2 \over c}}x\Big){\rm e}^{-x^2}
H_1(x)H_{2n+1}(x)dx .
\end{equation}
The definitions of the constants in the above expressions are 
\[
B=A+A', \ \ \
{\rho}={\widetilde{{u'}}}B^{-1}u, \ \ \ 
{\bar \rho}={\rho}-{1\over c} \gamma \gamma'.
\]
\[
R=3{\rm Tr}(B^{-1}A'\Lambda A),\ \ \
Q=2\,{\widetilde{{u'}}}B^{-1}A\Lambda A'B^{-1}u.
\]
\begin{equation}
\label{mebasic}
c^{-1}={\widetilde{{w^{(ij)}}}}B^{-1}w^{(ij)},\ \ \ \
\gamma=c\,{\widetilde{w^{(ij)}}}B^{-1}u,\ \ \ \gamma'=c\,
{\widetilde{w^{(ij)}}}B^{-1}u',
\end{equation}
where the $N\times N$ symmetric matrix $\Lambda$ is 
defined by $T-T_{\rm cm}=(1/2)\sum_{i, j}
\Lambda_{ij}{\bf p}_i\cdot {\bf p}_j$ and 
$w^{(ij)}$ is an $N\times 1$ one-column matrix defined by

\begin{equation} 
w^{(ij)}_k=\delta_{ki}-\delta_{kj} \ \ \ \ \ \ (k=1,...,N).
\end{equation}

\par\indent
The integral in Eq. (28) can be analytically evaluated for 
several potentials, including Coulomb, exponential or  Gaussian
potentials. The numerical evaluation for a general potential
is a simple matter and one tabulates  $I(c,n)$ for
the necessary values of $c$. 
For power law potentials $V(r)=r^k$, for example, including the
Coulomb interaction, the $c$-dependence of the integral 
$I(c,n)$ is factored out:
\begin{equation}
I(c,n)=\left({2 \over c}\right)^{k/2} I_n(k),
\end{equation}
where the remaining integral can be carried out and expressible in 
terms of the Gamma function: 
\begin{equation}
I_n(k)=\frac{1}{\sqrt{\pi}} \sum_{m=0}^{n} \frac{(-1)^m 2^{2n-2m+1}}
{m!(2n-2m+1)!}\Gamma \Big(n-m+\frac{k+3}{2}\Big).
\end{equation}
In particular, for the Coulomb force $(k=-1)$ we get 

\begin{equation}
I_n(-1)=\sqrt{\frac{4}{\pi}}\frac{(-1)^n}{(2n+1)n!}.
\end{equation}

The correlation function is calculated through the equation 
\[
\langle f'\vert \delta({\bf r}_i-{\bf r}_j-{\bf r}) 
\vert f \rangle= \langle f'\vert f\rangle 
c^{3\over 2}{\rm e}^{-{1\over 2}c r^2}
\left({\bar \rho \over \rho}\right)^L 
\sum_{n=0}^L   
r^{2n}\left({\gamma\gamma'\over \bar\rho}\right)^n 
\]
\begin{equation}
\qquad \times \, \sum_{\kappa=0}^n C_{Ln\kappa} 
\langle L M\, 2\kappa \, 0 \vert L M \rangle Y_{2\kappa \,0}({\hat {\bf r}}) ,
\end{equation}
where
\begin{equation}
C_{Ln\kappa}={(-1)^{\kappa}(2\kappa-1)!! \sqrt{(2L-2\kappa)!(2L+2\kappa+1)!}
\over \pi (2L-1)!! 2^{L+1/2} (L-n)! 
\kappa! (n-\kappa)! (2n+2\kappa+1)!!}\sqrt{{4\kappa+1\over 2L+1}}.
\end{equation}

\section*{Appendix B: Separation of center-of-mass motion}
The transformation between the relative and single-particle 
coordinates, given by Eqs. (\ref{eq:x1})-(\ref{eq:x4}), can 
be defined by the matrix:

\begin{equation}
U=
\left(
\begin{array}{cccc}
1 & -1 & 0 & 0  \\
0 &  0 & 1 & -1 \\
{M\over m+M} & {m\over m+M} & {-M\over m+M}  & {-m\over m+M} \\
{M\over 2m+2M} & {m\over 2m+2M} & {M\over 2m+2M}  & {m\over 2m+2M} 
\end{array} 
\right), \ \ \ \ 
U^{-1}=
\left(
\begin{array}{cccc}
{m\over m+M}  &  0 & {1\over 2} & 1  \\
{-M\over m+M} &  0 &  {1\over 2}& 1  \\
0 & {m\over m+M}   & -{1\over 2}& 1  \\
0 & {-M\over m+M}  & -{1\over 2}& 1  
\end{array} 
\right) .
\end{equation}
The transformation between the relative and the single-particle 
coordinates is given by
\begin{equation}
{\bf x}=U{\bf r}, \ \ \ \ \ 
{\bf r}=U^{-1}{\bf x} .
\end{equation}
Here ${\bf r}$ and ${\bf x}$ are column vectors containing $({\bf r}_1,...,
{\bf r}_4)$ and $({\bf x}_1,...,{\bf x}_4)$. 
By this transformation one can express the CG of the single-particle 
coordinates by the relative coordinates:
\begin{equation}
G_A({\bf r})={\rm exp} \lbrace -{1\over 2} \tilde{\bf r} A {\bf r}\rbrace 
={\rm exp} \lbrace -{1\over 2} \tilde{\bf x} {\bf A}  {\bf x}\rbrace 
\equiv G_{\bf A}({\bf x}),
\ \ \ \ 
{\bf A}=\widetilde{U^{-1}}AU^{-1}.
\end{equation}
The parameters ${\bf A}_{Ni}={\bf A}_{iN}$, $(i=1,...,N-1)$ 
connect the relative and 
center-of-mass variables, and give rise to an undesirable center-of-mass 
dependence of the wave function. To have a translational invariant 
basis, we require that
\[
{\bf A}_{Ni}=0, \ \ \ {\bf A}_{NN}=c,
\]
that is, 
\begin{equation}
\sum_{j=1}^N\sum_{k=1}^N A_{jk}U^{-1}_{ki}=0 
\ \ \ (i=1,...,N-1), \ \ \ \ \  \sum_{j=1}^N\sum_{k=1}^N A_{jk}=c,
\end{equation}
where $c$ is an arbitrary, positive constant common for each basis 
function. The second condition assures the finite norm of the basis function. 
By this requirement 
the relative and center-of-mass motion is separated in the exponential part 
of the basis function. 
\par\indent
To remove the center-of-mass contamination from the angular part, let us 
express the global vectors ${\bf v}$ in terms of relative coordinates:
\begin{equation}
{\bf v}=\sum_{i=1}^{N}u_{i} {\bf r}_i=\sum_{i=1}^N u_{i}\sum_{k=1}^{N}
U^{-1}_{ik} {\bf x}_k .
\end{equation}
This identity shows that by requiring
\begin{equation}
\sum_{i=1}^{N}u_{i} U^{-1}_{iN}=\sum_{i=1}^{N}u_{i}=0, 
\end{equation}
the global vector becomes translationally invariant. 
\par\indent
By fulfilling Eqs. (39) and (41) the basis is free from any problems with 
the center-of-mass 
motion. These conditions fix $N+1$ nonlinear parameters among $N(N+1)/2+N
=N(N+3)/2$ parameters. For $N=4$ there remain nine free parameters for each 
basis function. 

\section*{Appendix C: Sherman-Morrison formula}
As it is shown in Appendix A, the calculation of the matrix
elements requires the evaluation of the determinant and inverse
of the matrix $B$. In the SVM process we probe many random trials 
with different matrices. Let us assume that we change the matrix $A$ 
of nonlinear parameters in such a way that we change the parameter
$\alpha_{ij}\ (i\ne j)$ of the relative motion between particles $i$ and $j$
to $\alpha_{ij}+\lambda$ but keep all other matrix elements unchanged. 
This is certainly a very
restricted way, but in this case the computer time required for
the evaluation of the matrix elements tremendously decreases.
This change of $\alpha_{ij}$ produces the following changes in the matrix A 
(see Eq. (13)): 

\begin{equation}
A_{ij} \to A_{ij}-\lambda,\ \ \ \  A_{ji} \to A_{ji}-\lambda,\ \ \ \ 
A_{ii} \to A_{ii}+\lambda,\ \ \ \ A_{jj}\to A_{jj}+\lambda.
\end{equation}
It is easy to see that this change does not violates the conditions of 
Eq. (39). Thus the wave function with this modification is still 
translational invariant. 
The above change in the matrix $A$ can be simply expressed 
by using the vector $w^{(ij)}$ defined in Eq. (30) 
as follows: 
\begin{equation}
\label{changeofa}
A\rightarrow A - \lambda w^{(ij)}\widetilde{w^{(ij)}},
\end{equation}
Note that $w^{(ij)}\widetilde{w^{(ij)}}$ 
is an $N\times N$ matrix, whereas $\widetilde{w^{(ij)}}w^{(ij)}$ is 
just a number. As $B$ is equal to $A+A'$,
the above change leads to the following modification of $B$,
\begin{equation}
B\rightarrow B - \lambda w^{(ij)}\widetilde{w^{(ij)}}.
\label{eq:sher}
\end{equation}

To calculate the inverse and determinant of the above 
special form, the Sherman-Morrison formula can be used:
\begin{equation}
\left(B- \lambda w^{(ij)}\widetilde{w^{(ij)}}\right)^{-1}=
B^{-1}+ {\lambda  \over 1 - \lambda \widetilde{w^{(ij)}} B^{-1}w^{(ij)} }
B^{-1}w^{(ij)}\widetilde{w^{(ij)}}B^{-1} ,
\end{equation}
and
\begin{equation}
{\rm det}\left(B -\lambda w^{(ij)}\widetilde{w^{(ij)}}\right)=
\left(1- \lambda \widetilde{w^{(ij)}} B^{-1}w^{(ij)}\right) {\rm det}B .
\end{equation}
The advantage of this formulae is apparent: By knowing $B^{-1}$
and det$B$ one can easily calculate the right-hand sides of the equations,
and the $\lambda$ dependence is given in a very simple form. 
For example, $\widetilde{w^{(ij)}} B^{-1}w^{(ij)}$ simply reduces to 
$(B^{-1})_{ii}+(B^{-1})_{jj}-2(B^{-1})_{ij}$. Likewise, 
$B^{-1}w^{(ij)}\widetilde{w^{(ij)}}B^{-1}$ can also be easily evaluated. 
To change $\lambda$, therefore there is no need for the evaluation 
of inverses and determinants
(which would require $N^3$ operations) but we get the desired results by a 
simple multiplication and division. 

\section*{Appendix D: Symmetrization of wave functions}
\subsection*{Antisymmetrization}
The antisymmetrizer ${\cal A}$ is defined as
\begin{equation}
{\cal A}=\frac{1}{\sqrt{n_p}}\sum_{i=1}^{n_p} {\varepsilon_i} {\cal P}_i,
\end{equation}

\par\noindent
where the operator ${\cal P}_i$ changes the indices of identical
particles 
according to the permutation $(p_{1}^i,...p_{N}^i)$ of the numbers
$(1,2,...,N)$, and $\varepsilon_i$ is the phase of the permutation. The effect
of this operator on the set of the position vectors $({\bf r}_1,...,{\bf r}_N)$
is

\begin{equation}
{\cal P}_i ({\bf r}_1,...,{\bf r}_N)=({\bf r}_{p_1^i},...,{\bf r}_{p_N^i}).
\end{equation}

\par\noindent
By representing the permutations by the matrix

\begin{equation}
\left(C_i\right)_{kj}=1 \ \ \ \  {\rm if} \ \ \ \  j=p_{k}^i 
\ \ \ \ {\rm and}
\ \ \ \ \left(C_i\right)_{kj}=0 \ \ \ \ {\rm otherwise},
\end{equation}

\par\noindent
(for example, the permutation $(3\ 1\ 2\ 4)$
is represented by
\begin{eqnarray}
C=
\left(
\begin{array}{cccc}
0 & 0 & 1 & 0 \\
1 & 0 & 0 & 0 \\
0 & 1 & 0 & 0 \\
0 & 0 & 0 & 1 \\
\end{array}
\right),
\end{eqnarray}

\par\noindent
while for $(1\ 2\ 3\ 4)$ $C$ is a unit matrix),
the effect of the permutation operator on the single-particle coordinates 
reads as

\begin{equation}
{\cal P}_i {\bf r}=C_i{\bf r} .
\end{equation}

\par\noindent
By using  Eqs. (37) and (51) the permutation of the relative coordinates is
expressible as

\begin{equation}
{\cal P}_i {\bf x}=P_i{\bf x}\ \ \ \ \ {\rm with}\ \ \ \ \ \ P_i$=$UC_iU^{-1}.
\end{equation}

\par\noindent
The CGs, after permutation, take the form:

\begin{equation}
{\cal P}_i G_{A}({\bf r})=G_{\tilde{C_i}AC_i}({\bf r})=
G_{\tilde{P_i}{\bf A}P_i}({\bf x})=
G_{\widetilde{C_iU^{-1}}AC_iU^{-1}}({\bf x}).
\end{equation}

\par\indent
In the spin space the permutation operator interchanges the indices
of the single-particle spin functions and can be easily evaluated.
As a result, the matrix element of any spin-independent operator
$\cal O$ which is symmetrical with respect to the permutation of 
identical particle coordinates
can be written in the following form:
\begin{eqnarray}
& &\langle
{\cal A} \lbrace \chi_{SM_S} f_{KLM}(u',A',{\bf r}) \rbrace
\vert {\cal O} \vert
{\cal A}\lbrace \chi_{SM_S} f_{KLM}(u,A,{\bf r}) \rbrace
\rangle \nonumber \\
&=& \sum_{i=1}^{n_p} c_i\langle
f_{KLM}(u',A',{\bf r}) \vert {\cal O} \vert
f_{KLM}(\tilde{C_i}u,\tilde{C_i}AC_i,{\bf r}) \rangle ,
\label{mebasisfn}
\end{eqnarray}
where the coefficients $c_i$ have the form
\begin{equation}
c_i = \varepsilon_i \langle\chi_{SM_S}|{\cal P}_i|\chi_{SM_S}\rangle.    
\label{spincoef}
\end{equation}
Since the antisymmetrizer is a projector onto an antisymmetric state, 
only ket (or bra) function needs to be antisymmetrized. 

The particular value of the coefficient $c_i$ depends 
only on the spin function of the system. In the case of Ps$_2$ two 
positrons must be antisymmetrized and likewise two electrons must 
be in antisymmetric states. Therefore, the antisymmetrizer for this   
system is given by ${\cal A}=(1-P_{13})(1-P_{24})$, where $P_{ij}$ 
is the transposition of particle labels $i$ and $j$. Thus ${\cal A}$ 
has four permutations $(n_p=4)$ and we can identify 

\begin{equation}
{\cal P}_1=1,\ \ \ \ {\cal P}_2=P_{13},\ \ \ \ {\cal P}_3=P_{24},
\ \ \ \ {\cal P}_4=P_{13}P_{24}.
\end{equation}
The corresponding phases are $\varepsilon_1=1, \varepsilon_2=-1, 
\varepsilon_3=-1, \varepsilon_4=1$ and the matrices $C$ are given 
as follows:
 
\begin{equation}
C_1=
\left(
\begin{array}{cccc}
1 & 0 & 0 & 0 \\
0 & 1 & 0 & 0 \\
0 & 0 & 1 & 0 \\
0 & 0 & 0 & 1 \\
\end{array}
\right), \ \ \ 
C_2=
\left(
\begin{array}{cccc}
0 & 0 & 1 & 0 \\
0 & 1 & 0 & 0 \\
1 & 0 & 0 & 0 \\
0 & 0 & 0 & 1 \\
\end{array}
\right), \ \ \ 
C_3=
\left(
\begin{array}{cccc}
1 & 0 & 0 & 0 \\
0 & 0 & 0 & 1 \\
0 & 0 & 1 & 0 \\
0 & 1 & 0 & 0 \\
\end{array}
\right), \ \ \ 
C_4=
\left(
\begin{array}{cccc}
0 & 0 & 1 & 0 \\
0 & 0 & 0 & 1 \\
1 & 0 & 0 & 0 \\
0 & 1 & 0 & 0 \\
\end{array}
\right).
\end{equation}

The spin function 
$\chi_{00}$ of Eq. (\ref{spinzero}) is antisymmetric in both 
of the positron spin coordinates and the electron spin coordinates. 
Thus the spin matrix element $\langle\chi_{SM_S}|{\cal P}_i|
\chi_{SM_S}\rangle$ turns out to be equal to $\varepsilon_i$ and 
we have $c_1=c_2=c_3=c_4=1$.  

\subsection*{Charge symmetry}
The Hamiltonian $H$ for Ps$_2$ has charge-exchange symmetry, that is, 
it is invariant under the exchange of the positive and negative charges:
Letting $P$ denote the charge-permutation operator, we have

\begin{equation}
HP\psi=PH\psi=EP\psi.
\end{equation} 
Therefore, the non-degenerate eigenstate of the Hamiltonian is also 
the eigenstate of the charge-permutation operator.
In the Ps$_2$  the ground state is even $(\pi=+1)$ under $P$, 
while the $L=1$ excited state turns out to be odd $(\pi=-1)$. 

Consider the case of Ps $(e^+e^-)$.
This system is represented by the coordinate $({\bf r}_2 -{\bf r}_1)$.
The charge permutation is thus equivalent to the parity operation. 
Since the parity is $(-1)^L$ for the state with orbital angular momentum $L$, 
the eigenvalue of charge-permutation operator is also $(-1)^L$. 
The signs of ${\bf x}_1$ and $ {\bf x}_2 $ change with respect 
to the charge permutation $P_{12}P_{34}$, while ${\bf x}_3$ does not.
Assume that the Ps$_2$ has partial waves $l_1$, $l_2$ and $l_3$
corresponding to the motion described with 
${\bf x}_1$, $ {\bf x}_2 $ and $ {\bf x}_3 $, respectively.
When charges are permutated, the wave function $\psi$ become $(-1)^{l_1}
(-1)^{l_2}\psi$.
Then $P \psi = \psi$ for the $S$ state with $l_1 = l_2 =0$, while 
$P \psi = -\psi$ for the $P$ state with $l_1 = 0$ and $l_2 = 1$.

The non-vanishing matrix element of the electric dipole transition supports 
that the first excited $P$-state is odd under the charge permutation. 
This is because the electric dipole operator ${\bf D}$ has the following form 
except for the constant:

\begin{eqnarray}
{\bf D} = e({\bf r}_1-{\bf R}) -e({\bf r}_2-{\bf R}) 
    +e({\bf r}_3-{\bf R}) -e({\bf r}_4-{\bf R}),
\end{eqnarray}
which changes sign under the charge permutation. Therefore, if the excited 
$P$-state is even under the charge permutation, then the electric 
dipole matrix element between the $P$ state and 
the ground state would identically vanish. 

The charge-permutation operator $P$ 
is given by $P_{12}P_{34}$ or $P_{14}P_{32}$. 
When the wave function $\psi$ is already antisymmetrized for two positrons and 
for two electrons, then we can see that both operators give the same effect.
To understand this we use the following identity 

\begin{equation}
P_{14}P_{32}\psi=(P_{12}P_{34})^2 P_{14}P_{32} (P_{13}P_{24})^2 \psi
=P_{12}P_{34}P_{13}P_{24}\psi=P_{12}P_{34}\psi.
\end{equation}
Thus the basis function for the Ps$_2$ 
molecule with definite charge-permutation symmetry is given by 
operating with the following operator ${\cal C}$ on the function:

\begin{equation}
{\cal C}=\frac{1}{\sqrt{8}}(1+\pi P_{12}P_{34})(1-P_{13})(1-P_{24}).
\end{equation}

The evaluation of matrix elements between the 
states with odd charge symmetry can be done in a similar manner to the 
previous subsection by extending Eqs. (\ref{mebasisfn}) and 
(\ref{spincoef}). The antisymmetrizer ${\cal A}$ is now replaced with 
${\cal C}=({1}/{\sqrt{8}})\sum_{i=1}^8\varepsilon_i {\cal P}_i$, 
where new ${\cal P}_i$ are defined by 

\begin{equation}
{\cal P}_5=P_{12}P_{34},\ \ \  {\cal P}_6=P_{12}P_{34}P_{13},\ \ \  
{\cal P}_7=P_{12}P_{34}P_{24},\ \ \  {\cal P}_8=P_{12}P_{34}P_{13}P_{24},
\end{equation}
and the corresponding phases are $\varepsilon_5=-1$, $\varepsilon_6=1$, 
$\varepsilon_7=1$, $\varepsilon_8=-1$. The matrices $C_i$ corresponding 
to ${\cal P}_i$ are given below:
 
\begin{equation}
C_5=
\left(
\begin{array}{cccc}
0 & 1 & 0 & 0 \\
1 & 0 & 0 & 0 \\
0 & 0 & 0 & 1 \\
0 & 0 & 1 & 0 \\
\end{array}
\right), \ \ \ 
C_6=
\left(
\begin{array}{cccc}
0 & 0 & 0 & 1 \\
1 & 0 & 0 & 0 \\
0 & 1 & 0 & 0 \\
0 & 0 & 1 & 0 \\
\end{array}
\right), \ \ \ 
C_7=
\left(
\begin{array}{cccc}
0 & 1 & 0 & 0 \\
0 & 0 & 1 & 0 \\
0 & 0 & 0 & 1 \\
1 & 0 & 0 & 0 \\
\end{array}
\right), \ \ \ 
C_8=
\left(
\begin{array}{cccc}
0 & 0 & 0 & 1 \\
0 & 0 & 1 & 0 \\
0 & 1 & 0 & 0 \\
1 & 0 & 0 & 0 \\
\end{array}
\right).
\end{equation}
It is easy to evaluate the coefficients $c_i$. For the spin function 
$\chi_{00}$ of Eq. (\ref{spinzero}), we get $c_5=c_6=c_7=c_8=-1$.

\begin{table}
\caption{Comparision of the results of different calculations
for the ground-state energy of HPs. The energy is given in
atomic units.}
\begin{tabular}{lll}
Method& Reference & Energy     \\
\hline
SVM ($K=100$) &Present& $-$0.7891013600  \\
SVM ($K=200$) &Present& $-$0.7891810473  \\
SVM ($K=400$) &Present& $-$0.7891924458  \\
SVM ($K=800$) &Present& $-$0.7891958706  \\
SVM ($K=1200$)&Present& $-$0.7891964226  \\
SVM ($K=1600$)&Present& $-$0.7891965536  \\
Hylleraas configuration interaction & \protect\cite{clary}   & $-0.7842$              \\
Exponential trial functions         &\protect\cite{ho}  & $-0.7889$              \\
Diffusion Monte Carlo               & \protect\cite{yoshida}& $-0.7891\pm 0.002$     \\
Diffusion Monte Carlo               & \protect\cite{milano} & $-0.789175\pm 0.00001$ \\
Correlated Gaussian basis ($K=200$) & \protect\cite{posi3} & $-0.7891794$           \\
\end{tabular}
\end{table}

\clearpage

\begin{table}

\caption{Expectation values of various quantities for HPs. 
Atomic units are used. $K$ is the basis dimension.}
\begin{tabular}{lllll}
& $E$ & $-\langle V \rangle/(2 \langle T\rangle)$ \\
\hline
$K=100$& $-$0.7891013600& 1.00001  \\
$K=200$& $-$0.7891810473& 1.000003 \\ 
$K=400$& $-$0.7891924458& 1.000002 \\ 
$K=800$& $-$0.7891958706& 1.0000007\\ 
$K=1200$&$-$0.7891964226& 1.0000004\\
$K=1600$&$-$0.7891965536&1.0000003 \\
\hline
& $\langle r_{e^-e^-}^4 \rangle$ & $\langle r_{e^+e^-}^4 \rangle$ &
  $\langle r_{e^-p}^4 \rangle$ & $\langle r_{e^+p}^4 \rangle$ \\
\hline
$K=100$& 515.42669& 525.13203& 193.45055& 504.56556\\
$K=200$& 524.98363& 531.24425& 197.60909& 513.48089\\ 
$K=400$& 527.33506& 532.59188& 198.63996& 515.59169\\ 
$K=800$& 527.88970& 532.94707& 198.88278& 516.06972\\ 
$K=1200$&527.94660& 532.98328& 198.90610& 516.11702\\ 
$K=1600$&527.96159& 532.99639& 198.91176& 516.13646\\ 
\hline
& $\langle r_{e^-e^-}^3 \rangle$ & $\langle r_{e^+e^-}^3 \rangle$ &
  $\langle r_{e^-p}^3 \rangle$ & $\langle r_{e^+p}^3 \rangle$ \\
\hline
$K=100$& 83.599992& 83.792382& 34.789685& 84.226327\\
$K=200$& 84.337498& 84.249983& 35.120402& 84.911659\\
$K=400$& 84.507962& 84.347282& 35.195647& 85.064112\\
$K=800$& 84.544707& 84.369687& 35.211858& 85.094386\\
$K=1200$&84.548681& 84.372106& 35.213444& 85.097517\\
$K=1600$&84.549852& 84.372949& 35.213895& 85.098746\\
\hline
& $\langle r_{e^-e^-}^2 \rangle$ & $\langle r_{e^+e^-}^2 \rangle$ &
  $\langle r_{e^-p}^2 \rangle$ & $\langle r_{e^+p}^2 \rangle$ \\
\hline
$K=100$& 15.803193& 15.542251& 7.7797451& 16.188998\\
$K=200$& 15.860043& 15.575673& 7.8062352& 16.241186\\
$K=400$& 15.872464& 15.582575& 7.8117324& 16.252128\\
$K=800$& 15.874993& 15.584009& 7.8128668& 16.254178\\
$K=1200$&15.875286& 15.584176& 7.8129800& 16.254399\\
$K=1600$&15.875377& 15.584230& 7.8130152& 16.254480\\
\hline
& $\langle r_{e^-e^-} \rangle$ & $\langle r_{e^+e^-} \rangle$ &
  $\langle r_{e^-p} \rangle$ & $\langle r_{e^+p} \rangle$ \\
\hline
$K=100$& 3.5700072& 3.4777333& 2.3092381& 3.6573544\\
$K=200$& 3.5738023& 3.4797561& 2.3110943& 3.6607696\\
$K=400$& 3.5745993& 3.4801765& 2.3114423& 3.6614669\\
$K=800$& 3.5747568& 3.4802575& 2.3115152& 3.6616016\\
$K=1200$&3.5747763& 3.4802676& 2.3115221& 3.6616167\\
$K=1600$&3.5747825& 3.4802707& 2.3115245& 3.6616220\\
\hline
& $\langle r_{e^-e^-}^{-1} \rangle$ & $\langle r_{e^+e^-}^{-1} \rangle$ &
  $\langle r_{e^-p}^{-1} \rangle$ & $\langle r_{e^+p}^{-1} \rangle$ \\
\hline
$K=100$& 0.37072021& 0.41851818& 0.72973620& 0.34760250\\
$K=200$& 0.37058889& 0.41850815& 0.72971467& 0.34749891\\
$K=400$& 0.37056069& 0.41849668& 0.72970918& 0.34746907\\
$K=800$& 0.37055594& 0.41849614& 0.72970858& 0.34746293\\
$K=1200$&0.37055519& 0.41849601& 0.72970874& 0.34746209\\
$K=1600$&0.37055494& 0.41849596& 0.72970869& 0.34746180\\
\hline
& $\langle r_{e^-e^-}^{-2} \rangle$ & $\langle r_{e^+e^-}^{-2} \rangle$ &
  $\langle r_{e^-p}^{-2} \rangle$ & $\langle r_{e^+p}^{-2} \rangle$ \\
\hline
$K=100$& 0.21426165& 0.34877458& 1.2059515& 0.17234727\\
$K=200$& 0.21396622& 0.34911573& 1.2069510& 0.17221620\\
$K=400$& 0.21392019& 0.34912443& 1.2070112& 0.17217310\\
$K=800$& 0.21391300& 0.34914011& 1.2070561& 0.17216589\\
$K=1200$&0.21391137& 0.34914210& 1.2070629& 0.17216413\\
$K=1600$&0.21391064& 0.34914275& 1.2070632& 0.17216372\\
\hline
& $\langle{\bf r}\!_{e_a^-\!e_b^-}\!\cdot\!{\bf r}\!_{e_a^-\!e^+}\rangle$ &
  $\langle{\bf r}\!_{e^+\!e_a^-}\!\cdot\!{\bf r}\!_{e^+\!e_b^-}\rangle$ &
  $\langle{\bf r}\!_{pe_a^-}\!\cdot\!{\bf r}\!_{pe_b^-}\rangle$ &
  $\langle{\bf r}\!_{pe^+}\!\cdot\!{\bf r}\!_{pe^-}\rangle$ \\
\hline
$K=100$& 7.9015967& 7.6406546& $-$0.12185159& 4.2132458\\
$K=200$& 7.9300217& 7.6456510& $-$0.12378653& 4.2358745\\
$K=400$& 7.9362320& 7.6463425& $-$0.12449962& 4.2406428\\
$K=800$& 7.9374963& 7.6465132& $-$0.12462952& 4.2415176\\
$K=1200$&7.9376432& 7.6465328& $-$0.12466320& 4.2416014\\
$K=1600$&7.9376883& 7.6465421& $-$0.12467313& 4.2416325\\
\hline
& $-\langle \nabla_{e^-}^2\rangle$ &
  $-\langle \nabla_{e^+}^2\rangle$ &
  $\langle \nabla_{e_a^-}\cdot\nabla_{e_b^-}\rangle$ &
  $\langle \nabla_{e^+}\cdot\nabla_{e^-}\rangle$ \\
\hline
$K=100$& 0.65224870& 0.27367198& $-$0.043864431& 0.11701815\\
$K=200$& 0.65232846& 0.27369666& $-$0.043999455& 0.11707408\\
$K=400$& 0.65234077& 0.27369750& $-$0.044052593& 0.11707637\\
$K=800$& 0.65234481& 0.27369980& $-$0.044060768& 0.11707718\\
$K=1200$&0.652345728&0.27370016& $-$0.044063957& 0.11707760\\
$K=1600$&0.652345903&0.27370022& $-$0.044064366& 0.11707739\\
\hline
& $\langle \delta_{e^-e^-} \rangle$ & $\langle \delta_{e^+e^-}\rangle$ &
  $\langle \delta_{e^-p} \rangle$ & $\langle \delta_{e^+p}\rangle$ \\
\hline
$K=100$& 0.0047127& 0.0236658& 0.1717649& 0.0017964 \\
$K=200$& 0.0047873& 0.0242912& 0.1758767& 0.0016985 \\
$K=400$& 0.0044178& 0.0243887& 0.1761969& 0.0016542 \\
$K=800$& 0.0043895& 0.0244224& 0.1768711& 0.0016440 \\
$K=1200$&0.0043889& 0.0244583& 0.1771854& 0.0016386 \\
$K=1600$&0.0043867& 0.0244611& 0.1771862& 0.00163857\\
\end{tabular}

\end{table}

\begin{table}
\caption{Total energies of the Ps$_2$ molecule
in atomic units. $K$ is the basis dimension.}

\begin{tabular}{llc}\hline
Method  & Ps$_2$($L=0$) & Ps$_2$($L=1$) \\
\hline
SVM ($K=100$) & $-$0.516000069         &$-$0.334376975 \\
SVM ($K=200$) & $-$0.516003119         &$-$0.334405047 \\
SVM ($K=400$) & $-$0.516003666         &$-$0.334407561 \\
SVM ($K=800$) & $-$0.516003778         &$-$0.334408177 \\
SVM ($K=1200$)& $-$0.5160037869        &$-$0.334408234 \\
SVM ($K=1600$)& $-$0.516003789058      &$-$0.3344082658 \\
Ref.\cite{posi3} ($K=200$)& $-$0.5160024            &   \\
QMC  \cite{dario1}   & $-$0.51601$\pm$0.00001 &   
\end{tabular}
\end{table}

\clearpage
\begin{table}

\caption{Expectation values of various quantities for the ground state of 
Ps$_2$. Atomic units are used. 
The positrons are labeled 1 and 3 and the electrons are 2 and 4. 
Because of the charge-permutation symmetry, e.g. $\langle r_{12} \rangle=
\langle r_{14} \rangle=\langle r_{32} \rangle=\langle r_{34} \rangle$. 
$K$ is the basis dimension.}
\begin{tabular}{lllllll}\hline
& $\langle r_{13}^4 \rangle$ & $\langle r_{12}^4 \rangle$ &
  $\langle r_{13}^3 \rangle$ & $\langle r_{12}^3 \rangle$ &
  $\langle r_{13}^2 \rangle$ & $\langle r_{12}^2 \rangle$ \\
\hline
$K=100$&  5161.6174& 2786.7091& 442.51382& 252.36242& 46.328357& 29.088855\\
$K=200$&  5194.6167& 2803.5558& 443.64812& 252.94378& 46.368857& 29.109699\\
$K=400$&  5199.4736& 2805.9782& 443.77879& 253.00898& 46.372453& 29.111485\\
$K=800$&  5201.9725& 2807.2389& 443.85091& 253.04531& 46.374698& 29.112612\\
$K=1200$& 5201.9467& 2807.2264& 443.85059& 253.04519& 46.374696& 29.112613\\
$K=1600$& 5202.0371& 2807.2718& 443.85244& 253.04611& 46.374735& 29.112633\\
\hline
& $\langle r_{13}   \rangle$ & $\langle r_{12}   \rangle$ &
  $\langle r_{13}^{-1} \rangle$ & $\langle r_{12}^{-1} \rangle$ &
  $\langle r_{13}^{-2} \rangle$ & $\langle r_{12}^{-2} \rangle$ \\
\hline
$K=100$&6.0316960& 4.4863741& 0.22080676& 0.36840509& 0.073455963& 0.30308260\\
$K=200$&6.0330476& 4.4870759& 0.22079128& 0.36839678& 0.073445434& 0.30309811\\
$K=400$&6.0331385& 4.4871188& 0.22079076& 0.36839718& 0.073444789& 0.30310268\\
$K=800$&6.0332061& 4.4871525& 0.22079007& 0.36839692& 0.073444360& 0.30310349\\
$K=1200$&6.0332062&4.4871526& 0.22079008& 0.36839693& 0.073444319& 0.30310354\\
$K=1600$&6.0332070&4.4871530& 0.22079007& 0.36839693& 0.073444303& 0.30310361\\
\hline
& $\langle{\bf r}_{13}\!\cdot\!{\bf r}_{12}\rangle$ &
  $\langle{\bf r}_{12}\!\cdot\!{\bf r}_{14}\rangle$ &
  $\langle \delta(r_{13}) \rangle$ & $\langle \delta(r_{12})\rangle$ &
  $\langle \nabla_{1}\!\cdot\!\nabla_{2}\rangle$ &
  $\langle \nabla_{1}\!\cdot\!\nabla_{3}\rangle$ \\
\hline
$K=100$&23.164179& 5.9246760& 0.0006409& 0.0219092& 0.13077374& $-$0.00354409\\
$K=200$&23.184429& 5.9252702& 0.0006309& 0.0220330& 0.13077237& $-$0.00354402\\
$K=400$&23.186227& 5.9252581& 0.0006284& 0.0220860& 0.13077326& $-$0.00354475\\
$K=800$&23.186163& 5.9252654& 0.0006266& 0.0221064& 0.13077327& $-$0.00354466\\
$K=1200$&23.187348&5.9252652& 0.0006267& 0.0221075& 0.13077325& $-$0.00354461\\
$K=1600$&23.187368&5.9252651& 0.0006259&0.0221151&0.1307732538&$-$0.0035446132\\
\hline
& $\langle \nabla_{1}^2 \rangle$ & 
  $\vert 1+\langle V \rangle/(2 \langle T\rangle)\vert $\\\hline
$K=100$& $-$0.25800339& 0.7$\times 10^{-5}$\\
$K=200$& $-$0.25800073& 0.2$\times 10^{-5}$\\
$K=400$& $-$0.25800178& 0.1$\times 10^{-6}$\\
$K=800$& $-$0.25800188& 0.2$\times 10^{-7}$\\
$K=1200$&$-$0.25800188& 0.4$\times 10^{-8}$\\
$K=1600$&$-$0.258001894& 0.3$\times 10^{-9}$\\
\hline
\end{tabular}
\end{table}

\clearpage

\begin{table}

\caption{Expectation values of various quantities for the excited state of 
Ps$_2$. Atomic units are used. See the caption of Table IV.}
\begin{tabular}{lllllll}\hline
& $\langle r_{13}^4 \rangle$ & $\langle r_{12}^4 \rangle$ &
  $\langle r_{13}^3 \rangle$ & $\langle r_{12}^3 \rangle$ &
  $\langle r_{13}^2 \rangle$ & $\langle r_{12}^2 \rangle$ \\
\hline
$K=100$& 17822.007& 15534.005& 1222.7206& 1038.7198& 95.950622& 80.093853\\
$K=200$& 17925.902& 15603.238& 1226.3729& 1041.0599& 96.072859& 80.166513\\
$K=400$& 17937.861& 15611.357& 1226.7489& 1041.3065& 96.084420& 80.173591\\
$K=800$& 17939.361& 15612.015& 1226.7888& 1041.3221& 96.085316& 80.173768\\
$K=1200$&17939.589& 15612.121& 1226.7948& 1041.3249& 96.085461& 80.173821\\
$K=1600$&17939.574& 15612.112& 1226.7955& 1041.3251& 96.085514& 80.173836\\
\hline
& $\langle r_{13}   \rangle$ & $\langle r_{12}   \rangle$ &
  $\langle r_{13}^{-1} \rangle$ & $\langle r_{12}^{-1} \rangle$ &
  $\langle r_{13}^{-2} \rangle$ & $\langle r_{12}^{-2} \rangle$ \\
\hline
$K=100$&8.8538933& 7.5670069& 0.14726627& 0.24081436& 0.032251179& 0.16072903\\
$K=200$&8.8572758& 7.5686805& 0.14724521& 0.24082305& 0.032232174& 0.16080331\\
$K=400$&8.8575704& 7.5688316& 0.14724464& 0.24082544& 0.032230800& 0.16081241\\
$K=800$&8.8575804& 7.5688194& 0.14724481& 0.24082635& 0.032230213& 0.16081476\\
$K=1200$&8.8575826&7.5688189& 0.14724482& 0.24082644& 0.032230197& 0.16081489\\
$K=1600$&8.8575844&7.56881891&0.147244820&0.24082648& 0.032230158& 0.16081514\\
\hline
& $\langle{\bf r}_{13}\!\cdot\!{\bf r}_{12}\rangle$ &
  $\langle{\bf r}_{12}\!\cdot\!{\bf r}_{14}\rangle$ &
  $\langle \delta(r_{13}) \rangle$ & $\langle \delta(r_{12})\rangle$ &
  $\langle \nabla_{1}\!\cdot\!\nabla_{2}\rangle$ &
  $\langle \nabla_{1}\!\cdot\!\nabla_{3}\rangle$ \\
\hline
$K=100$& 47.975311& 32.118543& 0.0001590& 0.0108286& 0.09163822& $-$0.01610247\\
$K=200$& 48.036429& 32.130083& 0.0001509& 0.0111599& 0.09165330& $-$0.01610824\\
$K=400$& 48.042210& 32.131381& 0.0001482& 0.0111781& 0.09165593& $-$0.01610939\\
$K=800$& 48.042658& 32.131110& 0.0001463& 0.0112015& 0.09165677& $-$0.01610973\\
$K=1200$&48.042730& 32.131091& 0.00014627&0.0112016& 0.09165683& $-$0.01610972\\
$K=1600$&48.042757& 32.131079& 0.00014591&0.0112091& 0.091656853&$-$0.016109693\\
\hline
& $\langle \nabla_{1}^2 \rangle$ & 
  $\vert 1+\langle V \rangle/(2 \langle T\rangle)\vert $\\\hline
$K=100$& $-$0.1671740& 0.4$\times 10^{-4}$\\
$K=200$& $-$0.1671984& 0.1$\times 10^{-4}$\\
$K=400$& $-$0.1672025& 0.4$\times 10^{-5}$\\
$K=800$& $-$0.1672038& 0.8$\times 10^{-6}$\\
$K=1200$&$-$0.1672039& 0.5$\times 10^{-6}$\\
$K=1600$&$-$0.16720401&0.36$\times 10^{-6}$\\
\hline
\end{tabular}
\end{table}

\newpage

\begin{figure}
\bigskip
\begin{enumerate}

\epsfbox{HPs.epsf}
\item[ Fig. 1]: The correlation functions $r^2C({\bf r})$ for various 
pairs of the constituents of the hydrogen positride HPs. 
For the sake of comparison, the 
electron-proton correlation function of the H atom is also drawn.

\clearpage

\epsfbox{Ps2_L0.epsf}
\item[Fig. 2]:  The correlation functions $r^2C({\bf r})$ for the ground 
state of the Ps$_2$ molecule. The solid curve denotes the electron-electron 
correlation and the dashed curve the electron-positron correlation. For the 
sake of comparison, the electron-positron correlation function for the 
Ps atom is drawn by the dotted curve.

\clearpage

\epsfbox{Ps2_L1.epsf}
\item[Fig. 3]: The correlation functions $r^2C({\bf r})\ 
({\bf r}=(x,\,0,\,z))$, multiplied by one thousand, for the bound 
excited-state of the Ps$_2$ molecule which has the orbital angular 
momentum $L=1$, the spin $S=0$, and negative parity. The magnetic 
quantum number $M$ is set equal to 0 for (a) and to 1 for (b). 
Plotted on the $xz$ plane 
are the contour maps of the correlation function. 
\end{enumerate}

\end{figure}

\end{document}